\documentclass[twocolumn,journal]{IEEEtran}
\usepackage{graphicx}
\usepackage{amsmath}
\usepackage{amssymb}
\usepackage{cite}
\usepackage{booktabs}
\usepackage{dblfloatfix}

\title{Dynamic Channel Knowledge Map: Fundamentals, Construction, and Applications}
\author{Wenjun~Jiang,~\IEEEmembership{Member,~IEEE}, and Xiaojun Yuan,~\IEEEmembership{Fellow,~IEEE}}

\begin{document}

\bstctlcite{BSTcontrol}

\maketitle

\begin{abstract}
Wireless communication networks are evolving toward extremely large antenna arrays, millimeter-wave and terahertz bands, and dense heterogeneous deployments, all of which increase channel dimensionality and make channel acquisition increasingly costly. Channel knowledge map (CKM) establishes a mapping from geographical locations to channel characteristics, providing location-specific prior information to reduce the overhead of channel acquisition. Most existing CKM research, however, has focused on quasi-static propagation features shaped by quasi-static environmental structures such as buildings and terrain, leaving unaddressed the time-varying channel component introduced by dynamic scatterers, terminal attitude changes, and radio-frequency (RF) impairments. This article presents a new concept of dynamic CKM as a middle layer that links quasi-static environmental priors to physical-layer signal processing by providing time-evolving channel representations. We first introduce the fundamentals of dynamic CKM, clarifying its relationship with the quasi-static CKM and the physical layer. We then survey representative construction methods and discuss how dynamic CKM can support pilot design, interference suppression, and integrated sensing and communications. Finally, we outline key open research directions in the co-design of dynamic CKM construction and physical-layer signal processing. These discussions offer an architectural perspective on the role of dynamic CKM in emerging 6G systems.
\end{abstract}

\section{Introduction}

\begin{figure*}[!t]
\centering
\includegraphics[width=\textwidth]{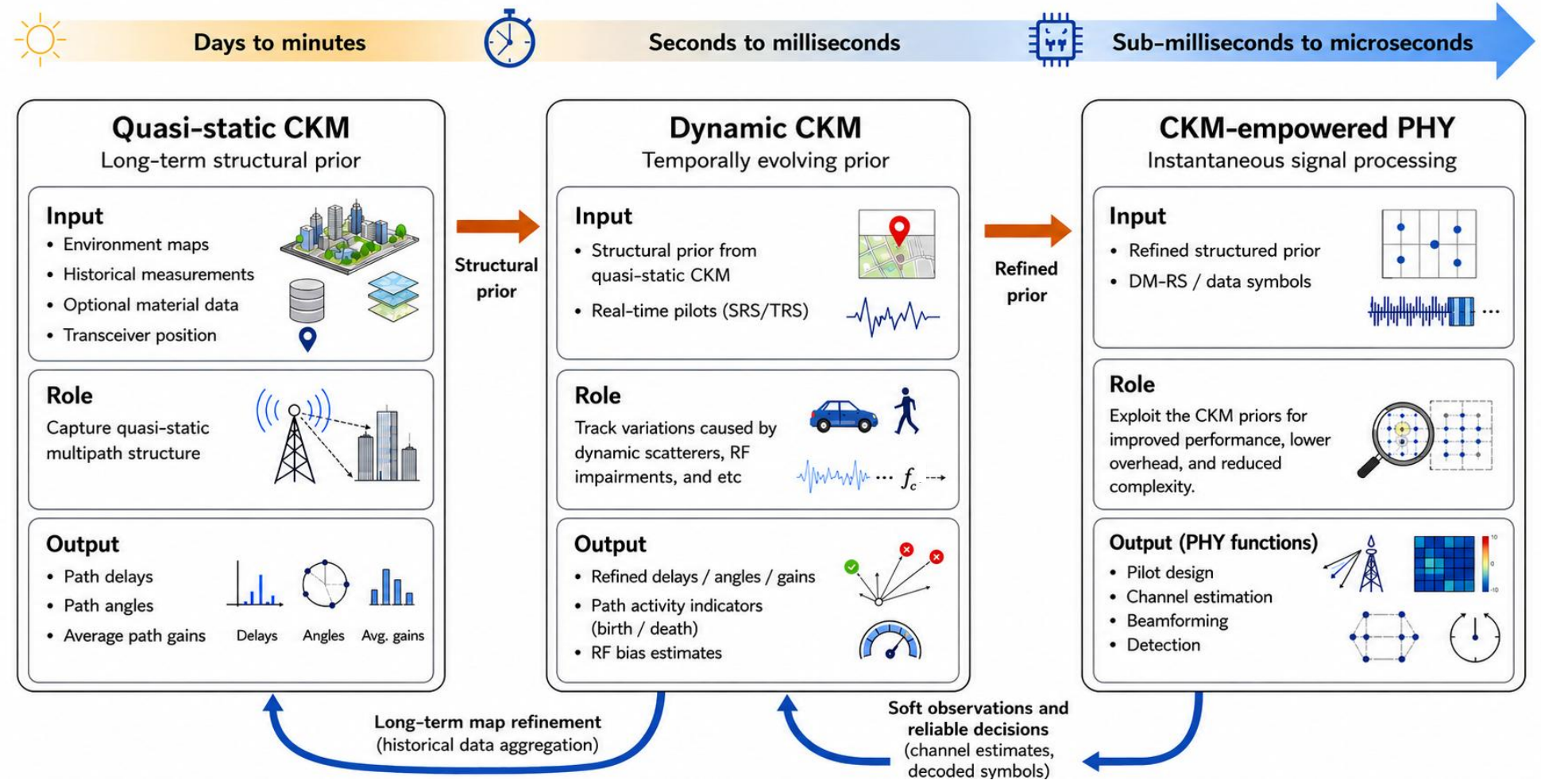}
\caption{Overview of the three-layer CKM framework. Quasi-static CKM captures quasi-static environmental propagation characteristics and provides structural priors to the dynamic CKM. Dynamic CKM tracks the dynamic channel component and delivers refined, temporally evolving priors to physical-layer signal processing. In the reverse information flow direction, physical-layer signal processing outputs channel estimates and reliable data-symbol decisions (as soft observations) that help refine the dynamic CKM.}
\label{fig:overview}
\end{figure*}

Wireless networks evolving toward 6G are driving wireless channels into a higher-dimensional and more dynamic regime. The deployment of extremely large antenna arrays (ELAAs) and the use of millimeter-wave (mmWave) and terahertz (THz) bands drastically enlarge the spatial and spectral dimensionality of the channel. Meanwhile, higher carrier frequencies introduce larger Doppler shifts, and emerging high-mobility scenarios, such as unmanned aerial vehicle (UAV) communications, further shorten the channel coherence time~\cite{sector2023framework}. This increase in channel dimensionality and temporal variation challenges physical-layer signal processing that relies on channel state information (CSI). CSI acquisition becomes increasingly costly because the number of pilot symbols scales with the channel dimensionality, and the pilot transmission interval is shortened as the channel coherence time decreases. Beam management in mmWave and THz systems is similarly affected, since rapid directional search must be conducted over much larger codebooks under tighter time constraints. In dense network deployments, inter-cell interference exhibits high-dimensional, fast-varying spatial structure that is difficult to characterize from pilot observations alone. These challenges call for a new CSI acquisition paradigm that can reduce the dependence on instantaneous pilot measurements.

Channel knowledge maps (CKMs) offer a promising way to address the above bottleneck. The central idea stems from a basic physical observation: given the boundary conditions of a propagation environment, including scatterer geometry, spatial layout, and material electromagnetic properties, the channel response between any transmitter and receiver pair can in principle be obtained based on Maxwell's equations. By constructing a mapping from geographical locations to multidimensional channel features, including path loss, multipath delays, angles of arrival/departure, and Doppler shifts, CKM transforms the channel acquisition paradigm from passive estimation without environmental awareness to active prediction informed by propagation priors~\cite{CKM_Zeng}. With the advancement of digital-twin and artificial-intelligence techniques, the concept of CKM is evolving toward a channel digital twin that builds a high-fidelity digital mirror of the physical propagation environment~\cite{DCT_Jianwen}.

Despite its promise, the vast majority of existing CKM research considers \emph{quasi-static} propagation environments. Quasi-static CKM captures channel parameters shaped by buildings, terrain, and other slowly varying structures. These channel parameters remain relatively stable over minutes to days and are well suited for base-station (BS) deployment planning, coverage evaluation, and resource scheduling. This stability, however, does not always hold. Vehicular movement, pedestrian motion, and terminal attitude variation can cause multipath parameters to vary on sub-second or even millisecond timescales. Radio-frequency (RF) impairments, such as oscillator drift and phase noise, further introduce time-varying non-idealities in the transceiver chain, changing the parameters of the effective channel. When a quasi-static CKM is applied in such dynamic conditions, the map prior may mismatch the instantaneous channel, thereby degrading physical-layer signal processing performance.

This article presents a new concept of \emph{dynamic CKM} as a middle layer between quasi-static CKM and real-time physical-layer signal processing. Dynamic CKM extends the map representation to include the time-varying channel component caused by mobile scatterers, antenna orientation changes, and RF non-idealities. The quasi-static map serves as an informative prior that enables efficient updating of these dynamic components. We first present the fundamentals of dynamic CKM and clarify its connections to quasi-static CKM and physical-layer signal processing in Section~II. We then survey representative construction methods and physical-layer applications in Section~III, where these two tasks are considered separately, and outline key open research directions in the co-design of dynamic CKM construction and physical-layer signal processing in Section~IV. In Section~V, we conclude the article with a summary and outlook.

\section{Fundamentals of Channel Knowledge Map}
\label{sec:fundamentals}

This section introduces the CKM framework and clarifies the timescales, data resources, and input/output boundaries of its three layers. Fig.~\ref{fig:overview} provides a schematic overview of the three-layer architecture.

\subsection{Three CKM Layers with Different Timescales and Data Resources}

\textbf{Quasi-static CKM} captures channel features determined by macroscopic environmental structures, including building geometry, terrain morphology, vegetation cover, and atmospheric conditions. Since these environmental factors usually change slowly, the quasi-static map typically updates on a timescale of minutes to days. The data resources for quasi-static CKM can include environment maps, such as 3D digital models, LiDAR point clouds, and satellite imagery, together with historical channel measurements and, when available, electromagnetic material databases. For a given transceiver position, the quasi-static map outputs a set of long-term stable multipath parameters, including path delays, angles of arrival and departure, and average path gains, which are largely determined by the propagation geometry. Representative construction methods include spatial interpolation, neural ray tracing, neural radiance fields, 3D Gaussian splatting, generative diffusion models, and message-passing-based methods~\cite{RT_NN,b4,NeRF_SS,6D_CKM,Xiucheng_Diffusion,Wenjun_CKM}.

\textbf{Dynamic CKM} tracks channel variations caused by mobile scatterers (e.g., vehicles and pedestrians), terminal attitude rotation, and time-varying RF impairments such as carrier frequency offset (CFO) and symbol timing offset (STO). These factors evolve much faster than changes in buildings and terrain, and the dynamic map typically updates on a timescale of milliseconds to seconds. When updating, the dynamic map uses the quasi-static CKM as a strong prior and can use reference signals, such as sounding reference signals (SRS) and tracking reference signals (TRS) standardized by 3GPP~\cite{TS38211}, whose transmission periods typically range from 5--160\,ms. Then, the dynamic CKM outputs refined quasi-static parameters, newly emerging dynamic multipath components, and calibrated RF error terms. Notably, these outputs do not include instantaneous multipath complex coefficients. These coefficients are highly sensitive to wavelength-scale changes in the propagation environment. Their estimation is therefore left to physical-layer signal processing.

\textbf{CKM-empowered physical layer} exploits the outputs of dynamic CKM for real-time tasks such as channel estimation, beamforming, signal detection, and pilot design. The channel parameters provided by dynamic CKM can, for instance, shrink the search space for channel estimation, assist beam optimization, and reduce pilot overhead. Such operations should complete within the time budget imposed by the standard frame  structure. In a typical 3GPP frame , $2$ pilot symbols, known as demodulation reference signals (DM-RSs), occupy the beginning of the frame  for channel estimation, and the remaining $12$ symbols are used for data detection. The corresponding frame  duration is typically $0.5$\,ms, and physical-layer processing operates on millisecond to sub-millisecond timescales.

The relationship among the three layers is summarized as follows. The quasi-static CKM provides a structural prior for the dynamic CKM, including multipath delays, angles, and average channel gains at a given transceiver location. Dynamic CKM then determines which quasi-static paths remain valid, which new dynamic paths appear, and what RF biases require calibration, thereby reducing prior mismatch at the physical layer. In this way, dynamic CKM delivers a refined and time-evolving prior for physical-layer processing. In the reverse direction, physical-layer signal processing feeds back channel estimates and reliable data-symbol decisions as soft observations, allowing the dynamic CKM to update its state without requiring additional pilot overhead.

\subsection{Why Dynamic CKM Is Necessary}

We demonstrate the necessity of dynamic CKM from two perspectives: the first is to show the impact of dynamic-channel estimation accuracy on physical-layer system performance, and the second is to clarify its distinctions from related concepts.

\begin{figure}
\centering
\includegraphics[width=\columnwidth]{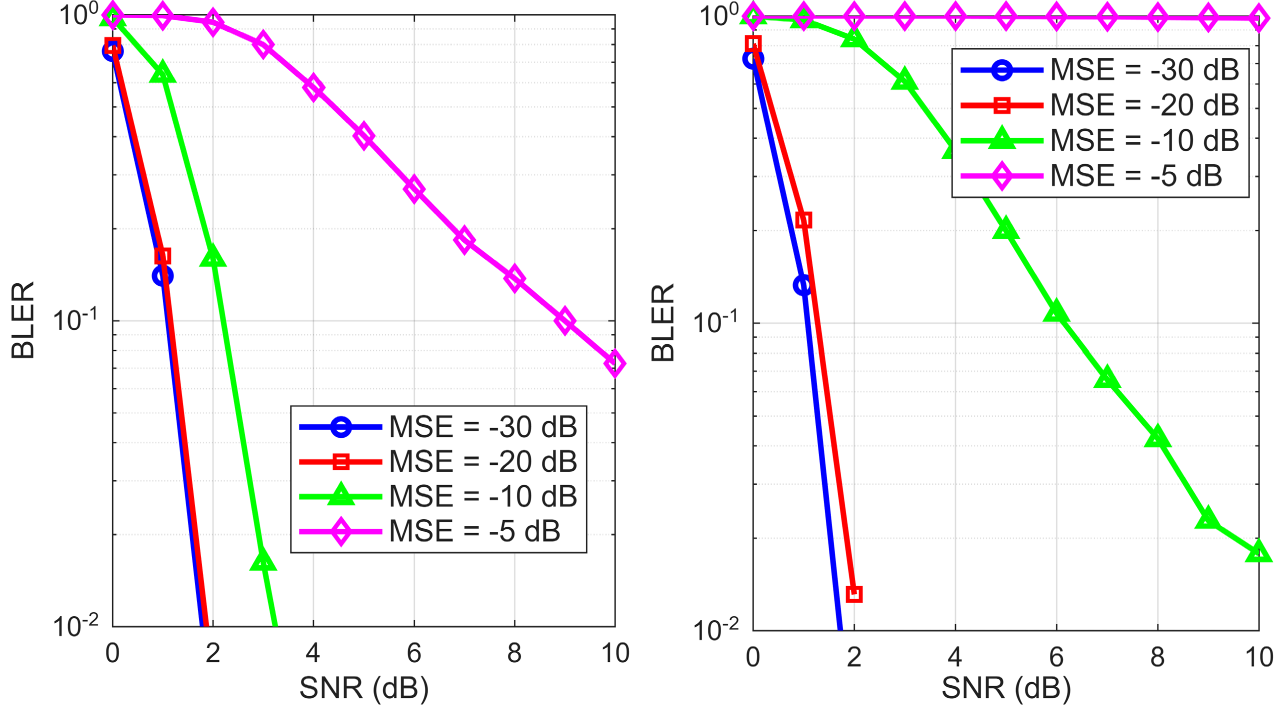}
\caption{Impact of dynamic-channel estimation accuracy on the block error rate (BLER) of a coded multiple-input multiple-output (MIMO) system. Left sub-plot: static-to-dynamic channel power ratio $E_s/E_d=10$~dB; Right sub-plot: $E_s/E_d=5$~dB. The system consists of a $64$-antenna BS serving $16$ uplink users and employs $64$-ary quadrature amplitude modulation (QAM) with low-density parity-check (LDPC) coding at a code rate of $0.85$. The quasi-static component is generated by the 3GPP TR~38.901 CDL-B model~\cite{TR38901} and assumed to be perfectly known, while the dynamic component follows a geometric-stochastic model with two clusters and $20$ subpaths per cluster.}
\label{fig:dynamic_impact}
\end{figure}

To start with, we consider a case in which channel dynamics arise from moving scatterers. Fig.~\ref{fig:dynamic_impact} shows how dynamic-component estimation accuracy affects coded MIMO detection performance. When the dynamic-to-static channel power ratio is $-10$~dB, accurate tracking of the dynamic component, with an estimation MSE of $-20$~dB or $-30$~dB, allows the receiver to reach $\mathrm{BLER}=0.1$ at a signal-to-noise ratio (SNR) of $1$~dB. Raising the estimation MSE to $-10$~dB shifts this operating point by roughly $1$~dB, whereas a further increase to $-5$~dB incurs an additional SNR loss of $8$~dB. When the dynamic-to-static channel power ratio rises to $-5$~dB, the penalty from estimation error intensifies. The BLER curve exhibits an error floor approaching unity for an estimation MSE of $-5$~dB, indicating that when the dynamic channel power is non-negligible, the estimation error will become the bottleneck of signal detection. This example reveals that moving scatterers constitute a major source of channel dynamics that affect physical-layer system performance when they are not accurately tracked. Besides, RF impairments and user attitude variations can also induce dynamic channel characteristics and merit explicit handling.

We now justify the necessity of dynamic CKM by showing its distinctions from two related concepts: dynamic radio map and temporal channel prediction. A \emph{dynamic radio map}~\cite{XiuCheng_Diff} updates a location-indexed channel metric, such as received signal strength (RSS) or path loss, without resolving individual multipath components. From this perspective, dynamic CKM can be regarded as a path-domain extension of the dynamic radio map, explicitly capturing the parameters of individual propagation paths. \emph{Temporal channel prediction} methods, such as Transformer architectures, forecast future channel states conditioned on historical CSI sequences. In existing practice, these historical sequences are directly mapped to a channel response, omitting further channel decomposition. By decomposing the channel response into quasi-static and dynamic components, dynamic CKM enables the temporal predictor to focus exclusively on the dynamic component and exploit environment-specific parameter priors, thereby improving prediction accuracy.

\subsection{Role of Dynamic CKM in the CKM Framework}

The preceding discussion positions dynamic CKM as an intermediate layer between quasi-static environmental priors and real-time physical-layer signal processing. Based on the quasi-static map, dynamic CKM calibrates the parameter deviations introduced by environmental dynamics and time-varying RF impairments, supplying updated multipath delays and angles, path activity status, and RF calibration parameters that static priors alone cannot provide.

As an intermediate layer, dynamic CKM supports two interaction modes with the physical layer. In the first mode, map updating and physical-layer processing remain separate: the dynamic CKM is updated first, and the physical layer then queries it as a lookup table. In the second mode, map updating and physical-layer signal processing algorithms are jointly designed so that the two modules can share observations, exchange intermediate estimates, and enhance each other. Sections~III and~IV describe these two modes in turn.

\section{Separate Dynamic CKM Construction and Physical-Layer Applications}
\label{sec:separate}

This section first surveys representative construction methods and then discusses how the constructed dynamic CKM can support physical-layer tasks.

\begin{table*}[t]
\centering
\caption{Comparison of Dynamic CKM Construction Methods}
\label{tab:construction}
\begin{tabular}{@{}lccc@{}}
\toprule
\textbf{Attribute} & \textbf{RadioDiff~\cite{XiuCheng_Diff}} & \textbf{RT-GSHCM~\cite{ChengXiang_Dynamic}} & \textbf{Approximate Bayesian Inference~\cite{Jiang2025DynCKM}} \\
\midrule
Dynamic scatterer positions & Assumed known & Not required & Not required \\
Terminal attitude & Not modeled & Not modeled & Modeled \\
RF impairments & Not modeled & Not modeled & Modeled (STO) \\
Core technique & Denoising diffusion model & RT and GBSM & Two-stage Bayesian inference \\
Prior information used & Environment map & Offline RT for quasi-static channel & Historical data collected from BS \\
Output & Grid-based path-loss & Grid-less channel parameters & Grid-less channel parameters \\
\bottomrule
\end{tabular}
\end{table*}

\subsection{Dynamic CKM Construction Methods}

Compared with quasi-static CKM, which has been extensively studied, research on dynamic CKM is still at an infancy stage. Existing works~\cite{XiuCheng_Diff,ChengXiang_Dynamic} focus on dynamic scatterers as the primary source of channel variations. When the dynamic-scatterer positions are available from external sensors such as LiDAR, radar, or traffic monitoring systems, RadioDiff~\cite{XiuCheng_Diff} uses these positions as conditioning inputs to a denoising diffusion model and generates dynamic path-loss maps. Although RadioDiff targets radio maps rather than CKM, it demonstrates a sensor-conditioned update mechanism in which external perception directly drives map generation. When such positions are not available, RT-GSHCM~\cite{ChengXiang_Dynamic} combines an offline ray-tracing (RT) model for quasi-static channel characterization with an online geometry-based stochastic channel model (GBSM) for dynamic channel modeling. In this framework, online measurements are used to extract multipath components, which are then compared with RT-generated components to update the stochastic parameters of dynamic clusters.

RadioDiff and RT-GSHCM address dynamic propagation effects due to moving scatterers. Moving scatterers, however, are not the only source of channel variations. Terminal antenna attitude changes and STO can also introduce channel dynamics. Specifically, terminal attitude rotation changes the effective antenna radiation pattern and hence the path-power distribution, while STO introduces a common group-delay shift across multipath components. To capture these factors, the authors in ~\cite{Jiang2025DynCKM} proposed a two-stage dynamic CKM construction method for MIMO orthogonal frequency-division multiplexing (MIMO-OFDM) systems based on approximate Bayesian inference. In Stage~I, quasi-static channel parameters are estimated from historical measurements together with STO calibration. In Stage~II, the estimated quasi-static parameters are used as informative priors for estimating dynamic channel parameters from limited real-time pilot observations.

Table~\ref{tab:construction} compares the aforementioned methods in terms of required side information, modeled dynamic factors, output granularity, and inference mechanism. RadioDiff assumes that dynamic-scatterer positions are available from external sensors and outputs grid-based path-loss maps that capture scatterer-induced channel variations. RT-GSHCM eliminates the need for scatterer position information by using a GBSM model update driven by online channel measurements, yielding grid-less channel parameters of dynamic clusters. The approximate Bayesian inference approach uses limited real-time pilot observations and extends the modeled dynamic factors to include terminal attitude and STO calibration.

\textbf{Open Challenges:} The aforementioned investigations validate the feasibility of dynamic CKM construction, while several open challenges remain. First, existing schemes typically assume that either the transmitter or the receiver remains attitude-stable. This assumption breaks down in non-terrestrial networks, where the serving node is itself an aerial or orbital platform with time-varying position and attitude. In UAV-to-user links, the two ends translate and rotate simultaneously, so that path delays, angles, and per-path gains vary; combined with rich ground scattering, this makes multipath energy and scatterer visibility change far faster than in the terrestrial single-side case. In satellite links, the orbital motion of the satellite together with user-terminal attitude variation drives rapid, large-scale changes in link geometry and Doppler. Both cases require the quasi-static to be indexed by the the 3D positions of both the transmitter and the receiver \cite{6D_CKM}. Then, the attitude-induced dynamic effects are then left to the dynamic CKM, which calibrates the per-path gain and effective angle variations that the position-indexed quasi-static prior cannot capture. This division also points to a second challenge: existing frameworks rely mainly on wireless measurements for such calibration, whereas these platforms expose native state information, namely onboard inertial measurement units and flight-control telemetry for UAVs and orbital ephemeris for satellites, which, together with radar and camera sensing, can be fused to assist dynamic CKM construction. How to exploit such multimodal information remains largely open.

\subsection{Dynamic-CKM-Empowered Physical-Layer Algorithms}

Once a dynamic CKM is constructed, its temporally evolving priors can enhance a range of physical-layer tasks. We highlight three directions in which dynamic CKM provides value beyond what the quasi-static CKM alone can provide.

\subsubsection{Integrated Sensing and Communications}

CKM inherently fuses environmental geometry with propagation information, giving it a natural role in integrated sensing and communications (ISAC). Specifically, quasi-static CKM enables non-line-of-sight (NLoS) localization through an inverse mapping. The inverse map stores multi-dimensional multipath signatures on a predetermined spatial grid, and a terminal's observed signature is matched against the CKM database to infer the terminal position~\cite{CKM_Zeng}. Dynamic CKM can further support ISAC by identifying multipath components that mismatch the quasi-static map. Such mismatched components can be attributed to previously unseen dynamic scatterers, allowing the BS to detect their presence and estimate their position directly from the uplink communication signal. This capability is enhanced when the BS is equipped with an ELAA. In that case, dynamic scatterers may lie in the near-field region, where spherical-wave propagation carries both angular and distance-dependent information, improving the resolution of location and velocity estimation.

\subsubsection{Channel Estimation and Pilot Optimization}

Bayesian channel estimators such as the minimum mean squared error (MMSE) estimator depend on channel covariance matrices, which are typically obtained by averaging over large spatial regions and long time intervals. A covariance matrix estimated in this way captures average propagation statistics but discards the location-specific and time-specific structure that dynamic CKM provides. By supplying the delay and angle support set and the birth/death status of dynamic paths, the dynamic map can compress the estimation search space beyond what a quasi-static prior alone can achieve. Studies in MIMO-OFDM systems suggest that a joint frequency-space-domain MMSE estimator exploiting dynamic CKM priors outperforms the estimator based on quasi-static CKM priors in a dynamic environment~\cite{Jiang2025DynCKM}. Dynamic CKM can also support differentiated pilot allocation across users. The map can assess propagation complexity and short-term channel variability along each user's geographical trajectory. Users in multipath-rich, rapidly changing environments may require denser pilots on subcarriers, whereas users in more stable conditions may tolerate fewer pilots or longer sounding intervals. Such adaptation can be formulated as a constrained resource-allocation problem that balances estimation accuracy and total pilot overhead~\cite{b6}.

\subsubsection{Resource Scheduling and Interference Suppression} 

In dense networks, resource scheduling based solely on instantaneous CSI feedback incurs substantial signaling overhead and computational cost. CKM mitigates this overhead by supporting two-timescale resource management. On the slow timescale, the quasi-static CKM guides long-term decisions such as user grouping, cooperative cluster partitioning, and candidate beam codebook generation. On the fast timescale, the dynamic CKM provides short-term corrections to these decisions. For example, when strong dynamic paths appear, the scheduler can adjust power allocation or beamforming directions; when two users are predicted to have highly correlated channels in the next scheduling period, the scheduler can avoid co-frequency reuse to suppress inter-user interference.

\section{Co-Design of Dynamic CKM Construction and Physical-Layer Algorithms}
\label{sec:synergy}

The preceding section treats dynamic CKM construction and physical-layer applications as sequential stages. This separation scheme does not fully exploit the fact that map updating and physical-layer inference rely on the same observations, operate on overlapping timescales, and can refine each other. Specifically, the dynamic map provides scenario-specific priors that assist channel estimation, signal detection, and beamforming. In return, channel estimates and detected data symbols can be fed back to refine the map. This mutual dependence motivates a co-design in which map updating and physical-layer processing form a closed loop rather than a one-way pipeline. The following subsections discuss three representative directions for this co-design.

\subsection{Joint Dynamic CKM Construction and User Localization}

\textbf{Co-Design Problem:}
Existing dynamic CKM construction methods generally assume that the user's position is known, for example, through a global navigation satellite system (GNSS). This position is used to query the quasi-static map and retrieve the quasi-static multipath parameters against which the dynamic channel component is separated and estimated. In indoor corridors, urban canyons, and underground environments, however, GNSS signals are weak or unavailable, and position information cannot be taken for granted. Without a reliable position, the BS cannot directly retrieve the correct quasi-static prior, and the dynamic channel component becomes difficult to identify. Therefore, a key co-design problem is how to perform dynamic CKM construction and user localization jointly when the position is unknown.

\textbf{Why Co-Design Matters:}
Localization and dynamic CKM construction are mutually dependent. Accurate positioning against the quasi-static CKM requires the dynamic channel components to be first separated out from the channel response; otherwise, dynamic paths will be misattributed to the quasi-static multipath predicted by the quasi-static CKM, yielding a systematic position bias. Conversely, accurate dynamic channel estimation relies on a reliable position, through which the quasi-static CKM can be queried to obtain the prior. This circular dependence motivates a joint inference framework in which the user position and the dynamic channel parameters are estimated simultaneously.

\textbf{Potential Solutions:}
Joint inference can be approached by treating the position as a latent variable whose likelihood is defined through the quasi-static map, enabling alternating refinement of position and dynamic-channel parameters. Data-driven alternatives, such as generative diffusion models, can further handle cases where the posterior is highly nonlinear or multimodal. In both cases, the key is to fully exploit the prior information offered by the quasi-static CKM together with the structural features of the dynamic component, such as channel sparsity and low-rankness. The main design challenges include the sensitivity to quasi-static map errors, the computational cost of iterative inference under real-time latency constraints, and the convergence analysis of alternating refinement.

\begin{figure}[h]
\centering
\includegraphics[width=\columnwidth]{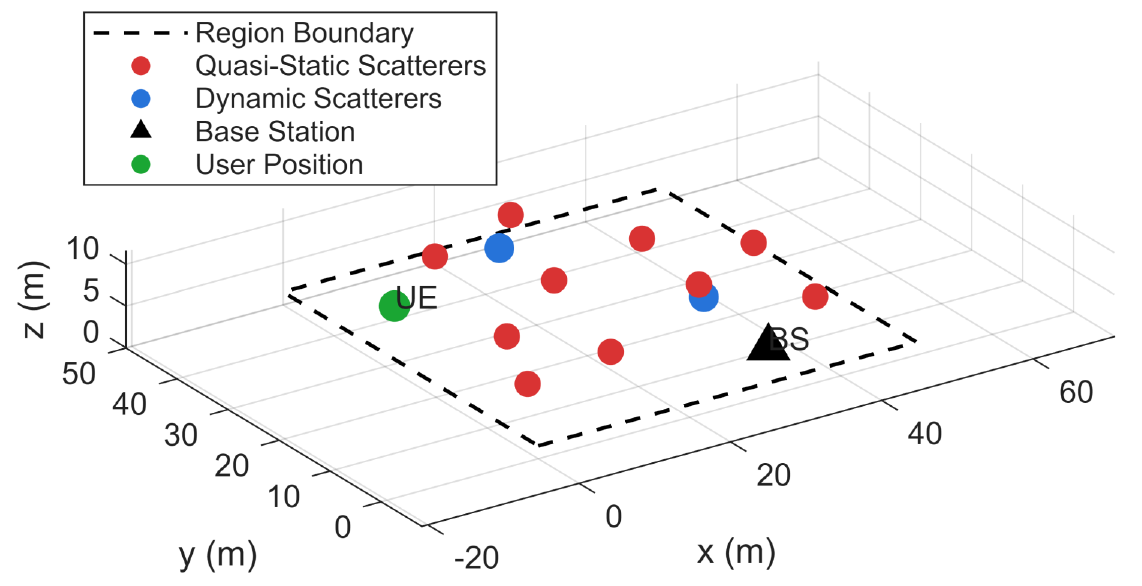}
\caption{A scenario used to illustrate the co-design of localization and dynamic CKM construction. A user moves in a $50\times 50$ m area, served by a BS equipped with an $8\times 16$ uniform planar array (UPA). The system uses $192$ pilot subcarriers spaced at $30$ kHz, and the quasi-static CKM is constructed on a $1$ m $\times$ $1$ m spatial grid. The static channel component is generated from a geometry-based channel model with $10$ quasi-static scatterers, while the dynamic component arises from $2$ dynamic scatterers.}
\label{fig:scenario}
\end{figure}

\begin{figure}[h]
    \centering
\includegraphics[width=0.85\columnwidth]{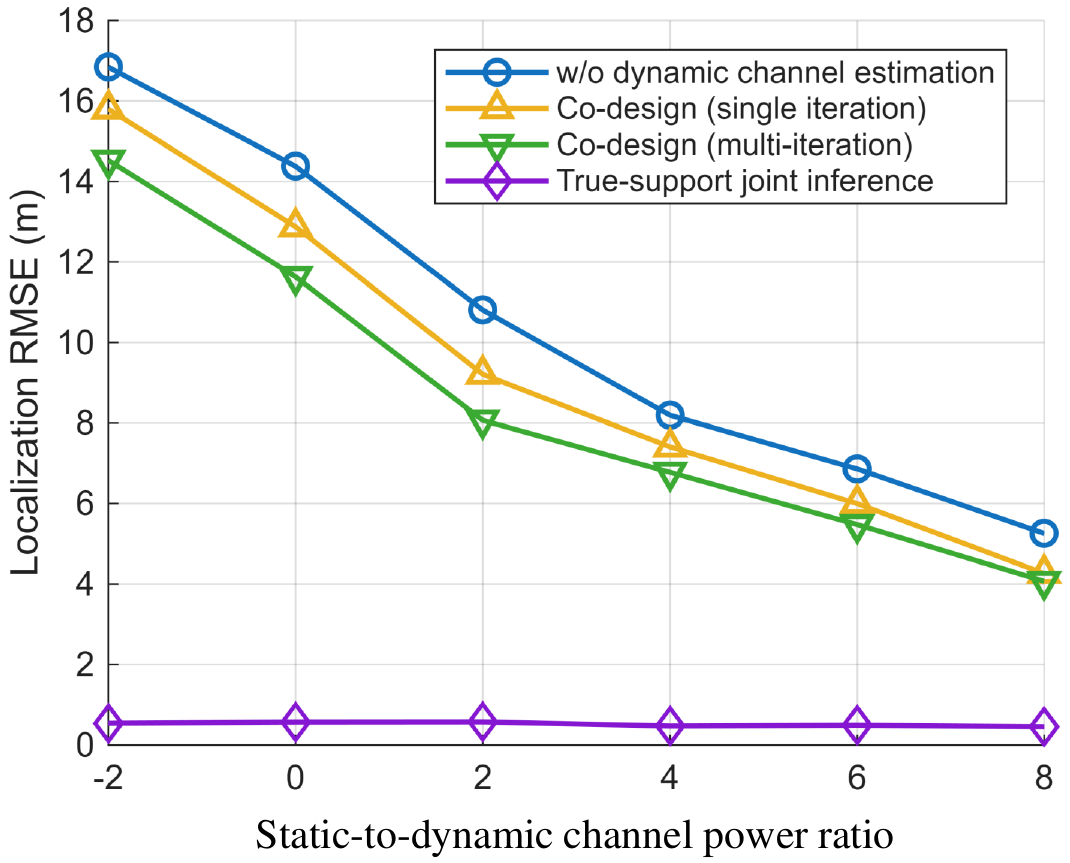}
    \caption{Localization root-mean-square error (RMSE) versus the static-to-dynamic power ratio $E_s/E_d$ at a SNR of $5$ dB. }
    \label{fig:ratio_rmse}
\end{figure}

For a preliminary evaluation of joint inference, we consider a scenario illustrated in Fig.~\ref{fig:scenario}. Correspondingly, Fig.~\ref{fig:ratio_rmse} shows the localization performance for four schemes: \emph{w/o dynamic channel estimation}, \emph{co-design (single iteration)}, \emph{co-design (multi-iteration)}, and \emph{true-support joint inference}. In the co-design receiver, a coarse position estimate is first obtained from the quasi-static CKM. The estimated quasi-static component is then subtracted to form a residual signal, from which the dynamic component is extracted. The position is finally re-estimated from the cleaned observation. The multi-iteration co-design repeats this loop. The true-support benchmark assumes that the dynamic delays and angles are perfectly known and performs joint inference of user location and dynamic channel response. The curves show that localization accuracy improves as the static component becomes more dominant. At $E_s/E_d=2$~dB, the multi-iteration co-design achieves a nearly $1.5$~m reduction relative to the single-iteration co-design. The gap between the co-design receivers and the true-support benchmark indicates that the current performance is still limited by simplified dynamic estimation and iterative updating. This gap suggests that localization-channel co-design remains an open problem that requires further investigation.

\subsection{Joint Dynamic CKM Construction and Channel Estimation}

\textbf{Co-Design Problem:}
In Section~III-B, dynamic CKM construction and channel estimation are treated as separate stages. The former tracks dynamic channel parameters using, e.g., SRS with a transmission period of 5--160\,ms, while the latter recovers the instantaneous channel response within a typical 0.5\,ms frame using DM-RS. The dynamic CKM outputs contain propagation channel parameters (e.g., dynamic scatterer-induced path delays and angles) and RF impairments (e.g., CFO and STO). Within one SRS sounding period, however, the dynamic channel parameters can undergo small drift. When these parameters are directly used without further refinement, the parameter drift will be translated into an irreducible channel estimation error. For example, the residual CFO introduces inter-carrier interference; the residual STO shifts the sampling instant and imposes a common group delay on both quasi-static and dynamic path delays. These residual RF impairments particularly affect high-throughput transmission, because the channel estimation error floor they induce will translate into an effective SNR ceiling that bounds the gains achievable from higher-order modulation and denser spatial multiplexing.

\textbf{Why Co-Design Matters:}
DM-RS pilots transmitted within every frame carry observations that, beyond supporting instantaneous channel estimation, also reflect the drift of dynamic channel parameters between SRS update intervals. Exploiting the DM-RS to refine the dynamic CKM can reduce the requirement on SRS resource and shorten the update interval of dynamic CKM. In return, a more frequently updated dynamic CKM supplies more accurate prior information on the dynamic channel parameters, against which DM-RS-based channel estimation can infer the multipath channel coefficients with higher accuracy. This bidirectional information flow motivates a joint design between dynamic CKM construction and channel estimation.

\textbf{Potential Solutions:}
From an algorithm-design perspective, we can develop a factor-graph-based inference algorithm, which uses message passing to iteratively estimate the dynamic channel parameters and the multipath channel coefficients~\cite{Wenjun_CKM}. The message-passing algorithm can be further unfolded into a neural network, where each iteration corresponds to a network layer and selected hyperparameters or submodules are optimized through data-driven training. The dynamic CKM prior is incorporated into each unfolded layer to assist the refinement of dynamic channel parameters and the estimation of multipath channel coefficients. The effectiveness of such inference rests on a fundamental theoretical question: under what conditions can the dynamic channel parameters and the multipath channel coefficients be resolved from finite DM-RS observations. Analyzing this separability condition would also guide pilot design. When multipath delays are poorly resolved from the available DM-RS subcarriers, increasing the pilot subcarrier density can improve the delay resolution; when the multipath channel coefficients vary rapidly across OFDM symbols, increasing the pilot symbol density can improve the temporal tracking.

\subsection{Joint Dynamic CKM Construction and Anti-Interference Signal Detection}

\textbf{Co-Design Problem:}
The above two subsections focus on the dynamic channel of target users. As cellular networks densify and sub-band full-duplex (SBFD) architectures are adopted~\cite{TS38211}, BSs encounter heterogeneous interference, e.g., co-link interference and cross-link interference. Specifically, co-link interference arises from uplink users in neighboring cells, whose signals propagate to the target BS via rich-scattering paths. Such interference exhibits moderate power and is broadly spread across the angular domain. In SBFD systems, cross-link interference arises from adjacent BSs transmitting downlink signals in the same sub-band. Since BSs are typically elevated and static, this interference is dominated by LoS or a small number of strong reflected paths, leading to high power concentrated within a narrow range of angular directions. Conventional methods based on power control and resource management fail to jointly mitigate these two structurally distinct types of interference.

\textbf{Why Co-Design Matters:} 
If the BS constructs a dynamic CKM that maps interferer locations to the corresponding propagation parameters, the BS can predict the interference subspace before data signal transmission and apply proactive interference suppression rather than reactive post-processing. Furthermore, the signal detector can produce recovered data symbols together with their reliability metrics. These outputs can serve as soft pilots for the estimation of interference channel parameters. In this way, the dynamic CKM for interferers can be updated without dedicating observation resources to individual interferers.

\textbf{Potential Solutions:}
A unified dynamic CKM framework would need to incorporate both co-link and cross-link interference channels. Under this framework, a hierarchical anti-interference scheme can be developed. Because co-link interference stems from neighboring uplink users and arrives via rich-scattering paths, it is generally weaker and more spatially diffusive. In this case, the dynamic interference map can provide the interference spatial covariance matrix, and a linear filter can use this covariance for subspace-based interference suppression. Cross-link interference stems from neighboring BSs and is typically stronger. In this case, subspace-based interference suppression alone can result in non-negligible energy leakage. A more suitable scheme may be interference reconstruction, where the interference CKM assists the detector in reconstructing the interference signal and canceling it from the observation before decoding the desired signal. From an analytical perspective, two fundamental questions merit investigation. The first is performance-complexity tradeoff, i.e., how the detection gain provided by a dynamic interference CKM scales with map accuracy and update period, so that the overhead of maintaining the map can be balanced against its benefit. The second is asymptotic performance analysis, i.e., under what conditions the hierarchical anti-interference scheme can approach the performance of an interference-free system.

\section{Conclusions and Outlook}
\label{sec:conclusions}

As wireless systems evolve toward ELAA, high-frequency bands, and dense heterogeneous deployments, the mismatch between the quasi-static CKM priors and the real-time channel dynamics becomes an increasingly important performance factor. This article has introduced dynamic CKM as a middle layer that operates on a millisecond-to-second timescale, tracking mobile scatterers, terminal attitude variations, and time-varying RF impairments while leveraging the quasi-static CKM as prior knowledge. We have reviewed initial efforts on dynamic CKM construction and discussed how dynamic priors can support channel estimation, beamforming, and ISAC. Beyond the construct-then-apply pipeline, we have also highlighted a co-design in which dynamic CKM construction and physical-layer processing can iteratively enhance each other.

Looking ahead, three directions may be particularly important in shaping the future development of dynamic CKM. The first is scalability. Low-complexity CKM construction methods are required to sustain real-time accuracy within the latency and energy constraints of large-scale deployments. The second is protocol design. Unified protocols for CKM representation, access, and update will be indispensable before dynamic CKM can be deployed in practical networks. The third is real-world validation. Field measurements and prototype experiments are needed to verify whether the gains achieved in simulation can be reproduced under realistic deployment conditions. As progress is made along these directions, dynamic CKM can evolve into an enabling infrastructure for environment-aware 6G physical-layer design.

\bibliographystyle{IEEEtran}
\bibliography{TurboMP}

\end{document}